\documentclass[
reprint,
nobibnotes,
 amsmath,amssymb,
 aps,
 pra,
]{revtex4-2}

\usepackage{graphicx}
\usepackage{dcolumn}
\usepackage{bm}
\usepackage{enumitem}
\usepackage{amssymb}
\usepackage{amsmath}
\usepackage{psfrag}
\usepackage{mathtools}
\usepackage{multirow}
\usepackage{color}
\usepackage{hyperref}
\usepackage[export]{adjustbox}
\usepackage{url}
\usepackage{stmaryrd}
\usepackage{dirtytalk}
\usepackage{cancel} 
\usepackage{braket} 
\usepackage{xcolor} 
\usepackage{enumitem}
\usepackage{cancel}
\usepackage{todonotes}

\usepackage{tikz} 

\definecolor{cyan1}{rgb}{0.387, 0.82, 1}
\definecolor{red1}{rgb}{0.902, 0.383, 0.355}
\definecolor{RGBred}{rgb}{1,0,0}
\definecolor{RGBblue}{rgb}{0,0,1}
\definecolor{outgreen}{rgb}{0.38,0.73,0.66}
\definecolor{wavered}{rgb}{0.933,0.196,0.18}
\definecolor{waveblue}{rgb}{0.133,0.337,0.651}
\definecolor{wallgrey}{rgb}{0.4,0.4,0.4}
\definecolor{weakblue}{rgb}{0.757,0.792,0.89}
\definecolor{weakred}{rgb}{0.973,0.745,0.745}
\definecolor{red2}{cmyk}{0,0.2,0,0}
\definecolor{fuchsia}{rgb}{0.57,0.36,0.51}
\definecolor{ao(english)}{rgb}{0.0, 0.5, 0.0}
\definecolor{amethyst}{rgb}{0.6, 0.4, 0.8}

\definecolor{green1}{cmyk}{0.85, 0.01, 1,0}

\newcommand{\cc}[1]{{#1}} 
\begin{abstract}
     \cc{
     We derive the analogues of the Dirac and Pauli equations from a spatially fourth-order Klein--Gordon equation with a universal length scale. Starting from a singularly perturbed variant of Maxwell's equations, we deduce a 32-dimensional variant of the Dirac equation for spin-$1/2$ particles through an algebraic factorization procedure. We illustrate an experimental test of the theory from the split lines of the electron beam in a Stern--Gerlach experiment. This hyperfine splitting leads to four distinct eigenvalues of the spin operator, which can be grouped into two pairs centered around the classic values of $\pm\hbar/2$. The modified electrodynamic framework features particle-antiparticle asymmetry and an oriented, micropolar spacetime.}
\end{abstract}    

\begin{document}

\title{\cc{Relativistic} electrodynamics with \cc{a universal} length scale}

\author{Tiemo Pedergnana}
\email{tiemop@mit.edu}
\affiliation{ Department of Mechanical and Process Engineering, ETH Zürich, Leonhardstrasse 21,  Zürich,
8092 Zürich, Switzerland}
\affiliation{
 Department of Mechanical Engineering, Massachusetts Institute of Technology, 77 Massachusetts Avenue, Cambridge, MA 02139, USA
}
\author{Florian Kogelbauer}
\affiliation{ Department of Mechanical and Process Engineering, ETH Zürich, Leonhardstrasse 21,  Zürich,
8092 Zürich, Switzerland}


\date{\today}%
\maketitle
\section{Introduction}
\subsection{Motivation: The Classical Klein--Gordon and Dirac Equations}


\cc{
The Klein--Gordon equation was the first successful attempt to reconcile quantum mechanics with special relativity, providing a relativistic wave equation for spin-0 particles by promoting the classical energy-momentum relation $E^2=c^2p^2+m^2c^4$ to the operator identity 
\begin{equation}
    (\hat{E}^2-c^2\hat{p}^2)\psi = m^2c^4 \psi,
\end{equation}
or, written out explicitly, 
\begin{equation}
\left(\nabla^2 - \frac{1}{c^2} \frac{\partial^2}{\partial t^2} \right) \psi = \frac{m^2 c^2}{\hbar^2} \psi, \label{Klein Gordon equation}
\end{equation}
where  $\psi=\psi(x,y,z,t)$ is a scalar wave function, $m$ is the particle mass and $c$ is the speed of light \cite{klein1926quantentheorie,gordon1926theorie}. 
The Klein--Gordon equation laid the groundwork for quantum field theory by showing how particle creation and annihilation arise naturally when one interprets $\psi$ as a field rather than a single-particle wavefunction. In modern physics, the Klein–Gordon equation underpins our understanding of mesons and other scalar fields \cite{bjorken1964relativistic}, serves as the prototype for constructing more complex gauge and spinor theories and illustrates fundamental concepts such as antiparticles, propagators, and the necessity of a field-operator formalism \cite{weinberg1995quantum}.\\
Although successful in the description of spin-$0$ particles, the second-order time derivatives and negative-energy solutions of the Klein--Gordon equation created interpretational difficulties, particularly in a single-particle context.\\
The Dirac equation \cite{Dirac1928,dirac1981principles}  for relativistic spin-$1/2$ particles overcame these issues by formulating a first-order differential equation in both space and time, naturally incorporating spin through the introduction of four-component spinor wavefunctions and gamma matrices.
Remarkably, the Dirac equation also predicted the existence of antiparticles, a profound insight that was later experimentally confirmed with the discovery of the positron \cite{anderson1933positive}. In this way, the Dirac equation not only extended the relativistic quantum framework but also complemented the Klein–Gordon approach, both of which are now seen as essential components in the development of quantum field theory, where particles of different spin are described by corresponding field equations.\\
Indeed, the Dirac equation is derived by a factorization of the Klein--Gordon equation by looking for an operator $\mathcal{D}$ and a wavefunction $\Psi$ satisfying 
\begin{eqnarray}
\mathcal{D}^2 \Psi &=&\left(\nabla^2 -\frac{1}{c^2} \frac{\partial^2}{\partial t^2} \right) \Psi \notag \\
&=&\frac{m^2 c^2}{\hbar^2} \Psi, \label{Factorization condition 2}
\end{eqnarray}
where $\hbar=h/(2\pi)$ is the reduced Planck constant. The Dirac equation is then given by the first-order system
\begin{eqnarray}
    \mathcal{D}\Psi=\dfrac{mc}{\hbar}\Psi. \label{Dirac equation}
\end{eqnarray}
In the non-relativistic limit of \eqref{Dirac equation}, Dirac famously recovered the Pauli equation for spin-1/2 particles \cite{Dirac1928,Pauli1927}. The dimensions of $\mathcal{D}$ and $\Psi$ along with their algebraic form are unknown \textit{a priori}. Dirac's fundamental insight was to assume $\Psi$ no longer as a scalar function, but much rather as a vectorial quantity. The operator $\mathcal{D}$ is then given by a 4-by-4 complex matrix differential acting on the four components of the vectorial wave function $\Psi$. The dimensionality of these quantities derives from algebraic constraints on the four Dirac matrices $\gamma_{i}$, $i=0,1,2,3$, which enter the picture through the following ansatz for $\mathcal{D}$:
\begin{eqnarray}
    \mathcal{D}=\gamma_0 \frac{i}{c} \frac{\partial}{\partial t} +\gamma_1 \frac{\partial}{\partial x}+\gamma_2 \frac{\partial}{\partial y} +\gamma_3 \frac{\partial}{\partial z}. \label{ansatz 1}
\end{eqnarray}
By substituting this expression into \eqref{Factorization condition 2}, one deduces that the following anti-commutator relations hold for the Dirac matrices \footnote{Depending on the chosen representation for the Dirac matrices, a signed diagonal matrix instead of the 4-by-4 unit matrix will be on the right-hand side of the following anti-commutator relation.}:
\begin{equation}
    \{\gamma_i,\gamma_j\}=2 \delta_{ij} \quad \forall\quad  i, j=0,\dots,3, \label{Anticomm. condition}
\end{equation}
where curled brackets denote the anti-commutator,
\begin{equation}
    \{a,b\}=ab+ba.
\end{equation}
}

\subsection{A Klein--Gordon equation with a universal length scale and broken Lorentz symmetry} 

\cc{
Special relativity considers the speed of light $c$ as an invariant in all inertial frames moving in uniform translation with respect to each other. As a consequence, any velocity can be expressed relative to $c$, while distance remains an absolute quantity. To remedy this asymmetry in phase space,  an additional universal length scale $\ell_0$ is introduced through the fourth-order equation proposed in  \cite{Pavlopoulos67}:
\begin{equation}
\left( -\ell_0^2 \nabla^4 + \nabla^2 - \frac{1}{c^2} \frac{\partial^2}{\partial t^2} \right) \psi = \frac{m^2 c^2}{\hbar^2}\psi. \label{Mindlin Pavlopoulos equation}
\end{equation}
For vanishing right-hand side, this equation was previously considered by \citet{Mindlin1962} in the context of micropolar plate mechanics and by \citet{Pavlopoulos67} as a nonlocal field theory with explicitly broken Lorentz symmetry. In particular, Pavlopoulos sought to tackle divergencies in present relativistic field theories, see also \cite{bopp1940,Feynman1948Classical,Feynman1948Quantum,Ji2019}, by replacing the d'Alembert operator 
\begin{eqnarray}
   \Box= \nabla^2 - \frac{1}{c^2} \frac{\partial^2}{\partial t^2},
\end{eqnarray}
appearing in the hyperbolic wave equation $\Box \psi=0$,  by the fourth-order operator
\begin{eqnarray}
    -\ell_0^2 \nabla^4 + \nabla^2 - \frac{1}{c^2} \frac{\partial^2}{\partial t^2}. \label{higher order operator}
\end{eqnarray}
The present work seeks to extend the analogy of Pavlopoulos by replacing $\Box$ in the Klein--Gordon equation \eqref{Klein Gordon equation}, and exploring the phenomena predicted by the resulting, modified wave equation.\\
The operator \eqref{higher order operator} studied in \cite{Pavlopoulos67} arises naturally from a variant of the vacuum Maxwell equations \cite{maxwell1865} under a singular, solenoidal (divergence-free) perturbation:
\begin{eqnarray}
  \nabla \cdot E &=&0, \label{max1}\\
    \nabla \cdot B &=& 0 \label{max2}, \\
    \nabla \times E &=& -\frac{\partial B}{\partial t},  \label{faraday}\\
    \nabla \times \left(B+\ell_0^2 \nabla \times \left(\nabla \times B\right)\right) &=& \frac{1}{c^2} \frac{\partial E}{\partial t} . \label{maxend}
\end{eqnarray}
Indeed, taking the partial time derivative of Eq. \eqref{faraday}, applying the curl to Eq. \eqref{maxend} and comparing the results yields 
\begin{eqnarray}
   \hspace{-0.3cm} \nabla\times \nabla\times B +\ell_0^2 \nabla \times\nabla \times \nabla\times \nabla\times B = -\frac{1}{c^2} \frac{\partial^2 B}{\partial t^2}.
   \label{eqyx} 
\end{eqnarray}
By the solenoidal nature of $B$ (see Eq. \eqref{max2}), Eq. \eqref{eqyx} is equivalent to 
\begin{eqnarray}
\left(-\ell_0^2 \nabla^2 \nabla^2+\Box\right)B=0. \label{B wave eq}
\end{eqnarray}
An analogous procedure, in which the partial time derivative of Eq. \eqref{maxend} is taken and the curl is applied to Eq. \eqref{faraday}, yields the same equation, but for $E$. No gauge conditions are required to arrive at these wave equations. In summary, the symmetries of Eq. \eqref{Mindlin Pavlopoulos equation} correspond to those of the modified electrodynamic framework given by Eqs. \eqref{max1} to \eqref{maxend}.
}


\cc{
The introduction of the length scale $\ell_0$ in the higher-order wave equation \eqref{Mindlin Pavlopoulos equation} breaks Lorentz symmetry. Indeed, the dispersion relation obtained in \cite{Pavlopoulos67} from setting the left-hand side of Eq. \eqref{Mindlin Pavlopoulos equation} equal to zero is given by 
\begin{eqnarray}
    \omega^2=c^2 k^2 (1+\ell_0^2 k^2).\label{dispersion}
\end{eqnarray}
The negative sign before the fourth-order term in Eq. \eqref{Mindlin Pavlopoulos equation} ensures that the frequency $\omega$ is strictly positive.
From \eqref{dispersion} we see immediately that any linear symmetry transform has to act independently on space and time, because $k$ and $\omega$ are of different leading order. Similarly, the two spatial differential operators of unequal order break dilation symmetry. Consequently, the classical linear action of the Lorentz group on equation \eqref{Mindlin Pavlopoulos equation} is reduced to Euclidean transformation of the spatial part.}

\subsection{Minimal coupling substitution}
{The minimal coupling substitution (see \cite[p. 205]{greiner2001quantum} and \cite[p. 409]{cohen2004photons}) is central to the derivation of the Pauli equation \cite{Pauli1927} from the Dirac equation \cite{Dirac1928,dirac1981principles}. The same is also true for the $\ell_0$-analogues of these equations, which are the focus of this study. For a particle with charge $q$, enforcing minimal coupling involves replacing the Hamiltonian $H$ and the impulse $p$ as follows:
\begin{eqnarray}
    H\rightarrow H+q\phi,\quad p\rightarrow \underbrace{p-q A}_{=\pi}, \label{minimal coupling}
\end{eqnarray}
where $\phi$ is the scalar electromagnetic potential,
\begin{equation}\label{defpi}
    \pi=p-q A,
\end{equation}
is the kinetic impulse and $A$ is the magnetic vector potential. Under this substitution, the dynamics of a charged particle with Hamiltonian $H= p^2/(2m)+V$, where $V$ is the potential, reproduce the Lorentz force law. Conversely, the Lorentz force law is equivalent to the Maxwell--Faraday equation combined with Faraday's law (see \cite[pp. 205--209]{landau1960electrodynamics}). Since Eqs. \eqref{max1} to \eqref{maxend} leave the Maxwell--Faraday equation (Eq. \eqref{faraday}) unchanged, the minimal coupling substitution \eqref{minimal coupling} reproduces the conventional Lorentz force law also in this modified framework of electrodynamics. This fact is used here to justify the application of the minimal coupling substitution \eqref{minimal coupling} in the context of the higher-order wave equation \eqref{Mindlin Pavlopoulos equation}.}


\section{Derivation of the Dirac equation with a universal length scale: Algebraic factorization}
The factorization method discussed above is now applied to Eq. \eqref{Mindlin Pavlopoulos equation}, which can be written out explicitly as
\begin{eqnarray}
&&\left( -\ell_0^2 \nabla^4 + \nabla^2 - \frac{1}{c^2} \frac{\partial^2}{\partial t^2} \right) \psi= \notag \\
&&\left[ -\ell_0^2 \left( \frac{\partial^4}{\partial x^4} + \frac{\partial^4}{\partial y^4} + \frac{\partial^4}{\partial z^4} + 2 \frac{\partial^4}{\partial x^2 \partial y^2}+ 2 \frac{\partial^4}{\partial x^2 \partial z^2}  \right.\right. \notag \\
&&\left.\left.+ 2 \frac{\partial^4}{\partial y^2 \partial z^2} \right)+ \frac{\partial^2}{\partial x^2} + \frac{\partial^2}{\partial y^2} + \frac{\partial^2}{\partial z^2}   -\frac{1}{c^2} \frac{\partial^2}{\partial t^2}\right]\psi \notag \\
&& =\frac{m^2 c^2}{\hbar^2} \psi.
\end{eqnarray}
To factorize this equation in the sense that 
\begin{eqnarray}
    \mathcal{D}_0^2 \Psi=\left( -\ell_0^2 \nabla^4 + \nabla^2 - \frac{1}{c^2} \frac{\partial^2}{\partial t^2} \right) \Psi, \label{requirement new}
\end{eqnarray}
in analogy to Eq. \eqref{ansatz 1}, the following ansatz is proposed here:
\begin{eqnarray}
\hspace{-0.2cm}\mathcal{D}_0&=& \Gamma_0 \frac{i}{c} \frac{\partial}{\partial t}+\Gamma_1 \frac{\partial}{\partial x}+\Gamma_2 \frac{\partial}{\partial y} +\Gamma_3 \frac{\partial}{\partial z} \notag\\
&&\left.+i \ell_0 \left[\Gamma_4\frac{\partial^2}{\partial x^2}+\Gamma_5\frac{\partial^2}{\partial y^2}+\Gamma_6\frac{\partial^2}{\partial z^2}\right. \right.\notag\\
&&\left.+{\sqrt{2}}\left(\Gamma_7\frac{\partial}{\partial x}\frac{\partial}{\partial y}+\Gamma_8\frac{\partial}{\partial x}\frac{\partial}{\partial z}+\Gamma_9\frac{\partial}{\partial y}\frac{\partial}{\partial z} \right)\right].\label{ansatz new}
\end{eqnarray}
By comparison of Eqs. \eqref{requirement new} and \eqref{ansatz new}, one can readily deduce that the $\Gamma$-matrices need to satisfy the anticommutation relations
\begin{equation}
    \{\Gamma_i,\Gamma_j\}=2 \delta_{ij} \quad \forall\quad  i, j=0,\dots,9. \label{Anticomm. condition new}
\end{equation}
These relations describe a Clifford algebra with $10$ elements, whose matrix representation has dimension $2^{10/2}=32$, see \cite{Lounesto2001}. An explicit construction of these matrices is due to \citet{weyl_brauer_1935}, and uses the Pauli matrices $\sigma_i$ ($i=1,2,3$):
\begin{eqnarray}
\hspace{-0.5cm}\sigma_1 &=& \begin{pmatrix}
0 & 1 \\
1 & 0
\end{pmatrix}, \quad
\sigma_2 = \begin{pmatrix}
0 & -i \\
i & 0
\end{pmatrix}, \quad
\sigma_3 = \begin{pmatrix}
1 & 0 \\
0 & -1
\end{pmatrix}.
\end{eqnarray}
The specific form of the Brauer--Weyl representation used here is slightly adapted from the version given in the original reference to ensure that the particle-antiparticle duality of the standard Dirac equation is also maintained for its $\ell_0$-analogue:
\cc{
\begin{eqnarray}
\Gamma_0 &=& \sigma_3\otimes I_2 \otimes I_2 \otimes I_2 \otimes I_2, \label{11}\\
\Gamma_1 &=& \sigma_2\otimes \sigma_3\otimes I_2 \otimes I_2 \otimes I_2, \\
\Gamma_2 &=& \sigma_2\otimes  \sigma_2\otimes \sigma_3\otimes I_2 \otimes I_2 ,\\
\Gamma_3 &=& \sigma_2\otimes  \sigma_2\otimes \sigma_2\otimes \sigma_3\otimes I_2 ,\\
\Gamma_4 &=& \sigma_2\otimes  \sigma_2\otimes   \sigma_2\otimes   \sigma_2\otimes  \sigma_3, \\
\Gamma_5 &=& \sigma_1\otimes I_2 \otimes I_2 \otimes I_2 \otimes I_2 ,\\
\Gamma_6 &=& \sigma_2\otimes  \sigma_1\otimes I_2\otimes I_2 \otimes I_2 ,\\
\Gamma_7 &=& \sigma_2\otimes   \sigma_2\otimes \sigma_1\otimes I_2 \otimes I_2, \\
\Gamma_8 &=& \sigma_2\otimes  \sigma_2\otimes \sigma_2\otimes \sigma_1\otimes I_2, \\
\Gamma_9 &=& \sigma_2\otimes  \sigma_2\otimes   \sigma_2\otimes   \sigma_2\otimes  \sigma_1,\label{12}
\end{eqnarray}
}
where $I_2$ is the two-by-two identity matrix. 

The $\Gamma$ matrices can be written as 
\begin{eqnarray}
    \Gamma_0=\begin{pmatrix}
        I_{16} &0\\
        0 &-I_{16}
    \end{pmatrix}, \label{block1}\\
    \Gamma_k=\begin{pmatrix}
        0 & G_k\\
        -G_k & 0
    \end{pmatrix}, \label{block2}\\
    \Gamma_5=\begin{pmatrix}
        0 & G_5\\
        G_5 & 0
    \end{pmatrix}, \label{block3}
\end{eqnarray}
for $k\in\{1, \dots, 9\} \setminus \{5\}$, where $G_k$ are 16-by-16 matrices given in the supplemental code \cite{supplement}, while $G_5=I_{16}$ is the $16$-by-$16$ identity matrix. Multiplication with $\Gamma_0$ changes the sign of the bottom half of the $\Gamma$ matrices:
\begin{eqnarray}
\Gamma_0\Gamma_k&=&\begin{pmatrix}
        0 & G_k\\
        G_k & 0
    \end{pmatrix},\quad k\in\{1, \dots, 9\} \setminus \{5\},\\
    \Gamma_0\Gamma_5&=&\begin{pmatrix}
        0 & G_5\\
        -G_5 & 0
    \end{pmatrix}. \label{eqsumxx}
\end{eqnarray}
Through the construction of \citet{weyl_brauer_1935} given by Eqs. \eqref{11} to \eqref{12}, there naturally emerges a central element 
$G_5$ which commutes with all other $G$-matrices: \begin{equation}
    \left[G_5,G_k\right]=0,\quad k=0,...,9.
\end{equation}
This commuting element leads to interesting physical consequences such as particle-antiparticle asymmetry and the emergence of a preferred direction in space, as shown in Secs. \ref{nonrel limit section} and \ref{example sec}. The choice of including $\Gamma_5$ in the $\ell_0$-modulated part of $\mathcal{D}_0$ is motivated by the correspondence principle.

The above steps lead to an explicit construction of the $32$-dimensional analogue of the Dirac equation corresponding to the modified Klein--Gordon equation \eqref{Mindlin Pavlopoulos equation}:
\begin{eqnarray}
    \mathcal{D}_0\Psi=\frac{m c}{\hbar}\Psi. \label{new Dirac}
\end{eqnarray}
In contrast to Eq. \eqref{Dirac equation}, $\Psi$ is now a $32$-dimensional wave function and $\mathcal{D}_0$ a $32$-by-$32$ complex matrix differential operator. After multiplying both sides with $c \hbar\Gamma_0$, Eq. \eqref{new Dirac} can be rearranged to yield 
\begin{eqnarray}
  i\hbar \frac{\partial}{\partial t} \Psi =\underbrace{\bigg\{ c p \cdot \left[\Gamma_0 \left(\Gamma_\text{s} +i\frac{\ell_0 }{\hbar} \mathcal{G} p \right)\right]+ mc^2 \Gamma_0 \bigg\}\Psi}_{=H\Psi},\label{simplified analogue dirac}
\end{eqnarray}
where $p=-i\hbar \nabla$ is the impulse operator and
\begin{eqnarray}
 \mathcal{G}&=&\frac{1}{{\sqrt{2}}}\begin{pmatrix}
{\sqrt{2}}\Gamma_4 & \Gamma_7 & \Gamma_8  \\
\Gamma_7 & {\sqrt{2}}\Gamma_5 & \Gamma_9 \\
\Gamma_8 & \Gamma_9 & {\sqrt{2}}\Gamma_6 
\end{pmatrix},\\
\Gamma_\text{s}&=-&i\begin{pmatrix}\Gamma_1 \\
\Gamma_2 \\
\Gamma_3
\end{pmatrix}.
\end{eqnarray}
The right-hand side of Eq. \eqref{simplified analogue dirac} is identified as the Hamiltonian $H$ acting on $\Psi$. After applying the minimal coupling substitution \eqref{minimal coupling}, Eq. \eqref{simplified analogue dirac} reads
\begin{eqnarray}
     i\hbar \frac{\partial}{\partial t} \Psi=\bigg\{ c \pi\cdot \left[\Gamma_0\left( \Gamma_\text{s} +i\frac{\ell_0}{\hbar}\mathcal{G} \pi \right)\right]+q \phi+ mc^2 \Gamma_0 \bigg\}\Psi, \notag \\
     \label{Symbolic 32d (min. coupling)}
\end{eqnarray}
where $\pi$ is the kinetic impulse \eqref{defpi}.\\
\cc{If all terms except the last inside the curled brackets are neglected, this equation reads
\begin{eqnarray}
     i\hbar \frac{\partial}{\partial t} \Psi= mc^2 \Gamma_0 \Psi. 
     \label{particle antiparticle form}
\end{eqnarray}
The $32$ eigenvalues of the energy operator on the left-hand side are $\pm mc^2$, each with $16$-fold degeneracy, corresponding to free particle and antiparticle solutions, respectively.\\
We note that, while the classical Dirac equation is invariant under spatial rotations \cite{dirac1981principles}, it is not immediately obvious that the rotational symmetry translates to the $l_0$-analogue \eqref{simplified analogue dirac}. Indeed, as already pointed out in \cite{Pavlopoulos67} and further elaborated in the discussion section of this paper, the full transformation group of \eqref{Mindlin Pavlopoulos equation} gives rise to a micropolar spacetime.}

\section{Nonrelativistic limit derivation \label{nonrel limit section}}

\subsection{Preliminaries}
In this section, the nonrelativistic limit of Eq. \eqref{Symbolic 32d (min. coupling)} is derived. To aid in this derivation, the following auxiliary quantities are first defined:
\begin{eqnarray}
    V &=&\begin{pmatrix}
        G_1 \\
        G_2 \\
        G_3 
    \end{pmatrix},\\
    M^\pm&=&\frac{1}{{\sqrt{2}}}\begin{pmatrix}
{\sqrt{2}}G_4 & G_7 & G_8  \\
G_7 & \pm{\sqrt{2}}G_5 & G_9  \\
G_8 & G_9 & {\sqrt{2}}G_6 
\end{pmatrix}, \label{eqsum}\\
M_0&=&\frac{1}{{\sqrt{2}}}\begin{pmatrix}
{\sqrt{2}}G_4 & G_7 & G_8  \\
G_7 & 0 & G_9  \\
G_8 & G_9 & {\sqrt{2}}G_6 
\end{pmatrix}. \label{eq9}
\end{eqnarray}
\cc{From the above definitions, it directly follows that 
\begin{eqnarray}
    M^+_{kl} V_m+ V_m  M^-_{kl}=0. \label{identity 2}
\end{eqnarray}
Contraction between two tensors $M$ and $N$ of equal order is defined as 
\begin{eqnarray}
    M:N=M_{ijkl\dots} N_{ijkl\dots}.
\end{eqnarray}
The Kronecker product between two tensors $P_{ij\dots}$ and $Q_{kl\dots}$ of arbitrary order is defined as 
\begin{eqnarray}
    [P\otimes Q]_{ij\dots kl\dots}=P_{ij\dots}Q_{kl\dots}.
\end{eqnarray}
If $P$ and $Q$ are vectors, then $P\otimes Q=PQ^T$.}

For the anticommutators of the $G$ matrices, the following relation holds:
\begin{eqnarray}
    \hspace{-0.2cm}\{G_j,G_k\}=-2\delta_{jk}+2(\delta_{j5}G_k+G_j\delta_{k5}). \label{gacom}
\end{eqnarray}
The spin vector $S$ appearing below is defined as follows:
\begin{eqnarray}
    \begin{pmatrix}
       S_1\\
        S_2\\
        S_3
    \end{pmatrix}&=&\frac{i\hbar}{4}\begin{pmatrix}
        G_2 G_3-G_3 G_2\\
        G_3 G_1-G_1 G_3\\
        G_1 G_2-G_2 G_1
    \end{pmatrix}. \label{seq}
\end{eqnarray}
Another quantity of interest is
\begin{eqnarray}
T_3=\frac{i\hbar}{4}\left(G_4 G_7-G_7 G_4\right). \label{T3 def}
\end{eqnarray}
The eigenvalues of $S_3-k^2\ell_0^2  T_3$, where $k$ is an arbitrary wavenumber, are given by 
\begin{eqnarray}
    \lambda_{++}&=&\frac{\hbar}{2}(1+k^2\ell_0^2),\label{ev1}\\
    \lambda_{+-}&=&\frac{\hbar}{2}(1-k^2\ell_0^2),\\
    \lambda_{-+}&=&-\frac{\hbar}{2}(1-k^2\ell_0^2),\\
    \lambda_{--}&=&-\frac{\hbar}{2}(1+k^2\ell_0^2). \label{ev4}
\end{eqnarray} 
Each of these eigenvalues corresponds to four linearly independent eigenvectors spanning the four-dimensional sectors $E_{++}$, $E_{+-}$, $E_{-+}$, and $E_{--}$. The matrix $S_3-k^2\ell_0^2 T_3$ thus has full rank and its eigenvectors are given in the supplemental code \cite{supplement}.\\

\subsection{Decoupling particle and antiparticle wave equations}
To obtain the non-relativistic limit of Eq. \eqref{Symbolic 32d (min. coupling)}, it is first rewritten in component form by making the substitution 
\begin{eqnarray}
    \Psi=e^{-i mc^2 t/\hbar }\begin{pmatrix}
    \xi\\
    \chi
\end{pmatrix}.
\end{eqnarray}
In these new variables, the left-hand-side of Eq. \eqref{Symbolic 32d (min. coupling)} reads
\begin{eqnarray}
i\hbar \frac{\partial}{\partial t}\Psi=e^{-i mc^2 t/\hbar } \left(mc^2 +i \hbar \frac{\partial}{\partial t} \right) \begin{pmatrix}
    \xi\\
    \chi
\end{pmatrix}, \label{LHS}
\end{eqnarray}
while the right-hand side reads
{\begin{widetext}
\begin{eqnarray}
   && \bigg\{ c \pi \cdot \left[\Gamma_0\left( \Gamma_\text{s} +i\frac{\ell_0}{\hbar} \mathcal{G} \pi \right)\right]+q \phi+ mc^2 \Gamma_0 \bigg\}\Psi  \notag\\
   &&= \left(c \pi \cdot \Biggl\{\Gamma_0\left[ -i\begin{pmatrix}\Gamma_1 \\
\Gamma_2 \\
\Gamma_3
\end{pmatrix} +i\frac{\ell_0}{{\sqrt{2}}\hbar}\begin{pmatrix}
{\sqrt{2}}\Gamma_4 & \Gamma_7 & \Gamma_8  \\
\Gamma_7 & {\sqrt{2}}\Gamma_5 & \Gamma_9 \\
\Gamma_8 & \Gamma_9 & {\sqrt{2}}\Gamma_6 
\end{pmatrix} \pi \right]\Biggr\}+q \phi+ mc^2 \Gamma_0\right) \Psi  \notag\\
 &&\cc{=\Biggl\{ -ic \pi \cdot \left[ \begin{pmatrix}\Gamma_0\Gamma_1 \\
\Gamma_0\Gamma_2 \\
\Gamma_0\Gamma_3
\end{pmatrix} -\frac{\ell_0}{{\sqrt{2}}\hbar}\begin{pmatrix}
{\sqrt{2}}\Gamma_0\Gamma_4 & \Gamma_0\Gamma_7 & \Gamma_0\Gamma_8  \\
\Gamma_0\Gamma_7 & {\sqrt{2}}\Gamma_0\Gamma_5 & \Gamma_0\Gamma_9 \\
\Gamma_0\Gamma_8 & \Gamma_0\Gamma_9 & {\sqrt{2}}\Gamma_0\Gamma_6 
\end{pmatrix} \pi \right]+q \phi+ mc^2 \Gamma_0 \Biggr\}\Psi } \notag\\
   && =e^{-imc^2 t/\hbar}\left[-ic\begin{pmatrix}
        0& \pi\cdot \left(V-\frac{\ell_0}{\hbar} M^+\pi \right) \\
        \pi\cdot \left( V -\frac{\ell_0}{\hbar} M^-\pi \right) & 0
    \end{pmatrix}+q\phi +mc^2 \begin{pmatrix}
        I_{16}& 0\\
        0 & -I_{16}
    \end{pmatrix} \right]\begin{pmatrix}
    \xi\\
    \chi
\end{pmatrix}. \label{RHS}
\end{eqnarray}
\end{widetext}
In arriving at this expression, the block properties of the $\Gamma$ matrices expressed in Eqs. \eqref{block1} to \eqref{eqsumxx} were exploited.} By comparing Eqs. \eqref{LHS} and \eqref{RHS} and rearranging (recall that $\xi$ and $\chi$ are $16$-dimensional wavefunctions), one obtains
\begin{eqnarray}
    \hspace{-0.3cm}i \hbar \frac{\partial}{\partial t} \xi &=&-ic\pi\cdot \left(V-\frac{\ell_0}{\hbar} M^+\pi \right) \chi+q\phi \xi, \label{eq1}\\
    \hspace{-0.3cm}i \hbar \frac{\partial}{\partial t} \chi&=&-ic\pi \cdot\left(V-\frac{\ell_0}{\hbar} M^-\pi \right) \xi+q\phi \chi-2mc^2\chi. \label{eq2}
\end{eqnarray}
In the nonrelativistic limit $c \rightarrow \infty$, both the $\hbar$-and $q$-modulated terms in Eq. \eqref{eq2} are small compared to the other terms, leading to
\begin{eqnarray}
    \chi\approx -i\left[\dfrac{\pi \cdot \left(V-\frac{\ell_0}{\hbar} M^-\pi\right)}{2mc}\right]\xi.
\end{eqnarray}
Substituting this result back into Eq. \eqref{eq1} yields
\begin{eqnarray}
    i \hbar \frac{\partial \xi}{\partial t}&=&- \dfrac{\left[\pi \cdot \left(V-\frac{\ell_0}{\hbar}M^+\pi\right)\right]\left[\pi \cdot \left(V-\frac{\ell_0}{\hbar}M^-\pi\right)\right]}{2m}\xi \notag \\
    &&+q\phi \xi.\label{Pauli Eq. before manipulation}
\end{eqnarray}
The first term on the right-hand side of this 16-dimensional equation is now examined, noting that the entries of $\pi$ commute with those of $V$ and $M^\pm$:
\begin{eqnarray}
   &&\left[ \pi \cdot \left(V-\frac{\ell_0}{\hbar} M^+\pi\right)\right] \left[ \pi \cdot \left(V-\frac{\ell_0}{\hbar} M^-\pi\right)\right] \nonumber\\
   &=&\pi_k\left(V_k-\frac{\ell_0}{\hbar} M^+_{kl}\pi_l\right) \pi_m\left(V_m-\frac{\ell_0}{\hbar} M^-_{mn}\pi_n\right) \notag \\
   &=&\underbrace{\pi_k \pi_m V_k  V_m}_{=\mathrm{I}}-\underbrace{\frac{\ell_0}{\hbar} \pi_k  \left(M^+_{kl}\pi_l\pi_m V_m+ V_k \pi_m M^-_{mn}\pi_n\right) 
}_{=\mathrm{II}}\notag\\
   &&+\underbrace{\left(\frac{\ell_0}{\hbar}\right)^2\pi_k\pi_l \pi_m \pi_n M^+_{kl} M^-_{mn}}_{=\mathrm{III}}. \label{Different terms}
\end{eqnarray}
{Using the above definitions, Eq. \eqref{Pauli Eq. before manipulation} can be restated as follows:
\begin{eqnarray}
    i \hbar \frac{\partial \xi}{\partial t}&=&- \dfrac{\mathrm{I}-\mathrm{II}+\mathrm{III}}{2m}\xi+q\phi \xi.\label{Pauli Eq. before manipulation, I terms}
\end{eqnarray}}

\subsection*{I}
The different terms in Eq. \eqref{Pauli Eq. before manipulation, I terms} are now separately examined. Let us begin with $\mathrm{I}$:
\begin{eqnarray}
   \mathrm{I}&=&\pi_k \pi_m V_k V_m \notag \\
   &=&\frac{\pi_k \pi_m}{2} \left[\left( V_k V_m -V_m V_k\right)+\left( V_k V_m +V_m V_k\right)\right]\notag \\
    &=&\frac{1}{4}\{ \pi_k, \pi_m\} (V_k V_m+V_m V_k)\notag \\
    &&+\frac{1}{4}\left[\pi_k ,\pi_m\right]\left( V_k V_m -V_m V_k\right),  \label{eq3}
\end{eqnarray}
where squared brackets denote the commutator,
\begin{eqnarray}
    [a,b]=ab-ba. 
\end{eqnarray}

Using the definition of the kinetic momentum operator $\pi$, noting that \cc{$\left[p_k,p_m\right]=0$ and $\left[A_k,A_m\right]=0$,} one can simplify the commutator in the above expression to 
\begin{eqnarray}
    \left[\pi_k ,\pi_m\right]&=&\left(p_k-qA_k\right)\left(p_m-qA_m\right)\notag \\
    &&-\left(p_m-qA_m\right)\left(p_k-qA_k\right)\notag \\
    &=&-q \left(p_k A_m +A_k p_m - p_m A_k -A_m p_k \right)\notag \\
    &=&-q\left(\left[p_k, A_m\right]-\left[p_m,A_k\right]\right)\notag \\
    &=&i\hbar q\left(\frac{\partial A_m}{\partial x_k} -\frac{\partial A_k}{\partial x_m}\right). \label{smth}
\end{eqnarray}
The term in the brackets represents the entries of a skew-symmetric matrix whose vector representation is the magnetic field $B$:
\begin{eqnarray}
 \frac{\partial A_m}{\partial x_k} -\frac{\partial A_k}{\partial x_m}=\varepsilon_{kml} B_l,
\end{eqnarray}
where $\varepsilon$ is the Levi-Civita symbol. Combining this result with Eq. \eqref{smth} yields
\begin{eqnarray}
    \left[\pi_k ,\pi_m\right]=i\hbar q\varepsilon_{kml} B_l.  \label{magB}
\end{eqnarray}

To make an analogy to the classic Pauli equation, the skew-symmetric 16-by-16 matrices $\Sigma_l$ and $S_l$, $l=1,2,3$, are here defined via (see also Eq. \eqref{seq}) 
\begin{eqnarray}
   (V_k V_m-V_m V_k)&=&2i\varepsilon_{kml} \Sigma_l, \label{exy}\\
   S&=&-\frac{\hbar}{2}\Sigma, \label{Seq}
\end{eqnarray}
where $l=1,\dots,3$. The vector $\Sigma$ is now written out explicitly by multiplying both sides of \eqref{exy} by $\varepsilon_{kmj}$:
\begin{eqnarray}
      \varepsilon_{kmj} (V_k V_m-V_m V_k)&=&2i\underbrace{\varepsilon_{kmj}\varepsilon_{kml} }_{=2\delta_{jl}}\Sigma_l, \notag\\
       \varepsilon_{kmj} (V_k V_m-V_m V_k)&=&4i \delta_{jl}\Sigma_l,  \notag\\
 2V_k V_m \varepsilon_{kmj}&=&4i \Sigma_j.\\
 \Rightarrow\Sigma_j&=&-\frac{i}{2} V_k V_m \varepsilon_{kmj}.
\end{eqnarray}
Therefore, 
\begin{eqnarray}
    \begin{pmatrix}
       \Sigma_1\\
        \Sigma_2\\
        \Sigma_3
    \end{pmatrix}&=&-\frac{i}{2}\begin{pmatrix}
        G_2 G_3-G_3 G_2\\
        G_3 G_1-G_1 G_3\\
        G_1 G_2-G_2 G_1
    \end{pmatrix}.
\end{eqnarray}
Finally, from Eq. \eqref{gacom}, one can deduce that 
\begin{eqnarray}
    (V_k V_m+V_m V_k)=-2\delta_{km}. \label{thiseq}
\end{eqnarray}
Combining Eqs. \eqref{eq3} to \eqref{thiseq} yields
\begin{eqnarray}
    \mathrm{I}&=&
    \pi_k \pi_m V_k  V_m\notag \\
    &=&\frac{1}{4}\{ \pi_k, \pi_m\} (V_k V_m+V_m V_k)\notag \\
    &&+\frac{1}{4}\left[\pi_k ,\pi_m\right]\left( V_k V_m -V_m V_k\right) \notag \\
    &=&-\frac{1}{2}\{ \pi_k, \pi_m\} \delta_{km}+\frac{i\hbar q}{4}\varepsilon_{kml} B_l\left( V_k V_m -V_m V_k\right)\notag \\
    &=&-\frac{1}{2}\{ \pi_k, \pi_m\} \delta_{km}-\frac{\hbar q}{2} \varepsilon_{kml}B_l \varepsilon_{kmj} \Sigma_j \notag \\
    &=&-\pi_k \pi_k-\hbar q \delta_{lj}B_l \Sigma_j \notag \\
    &=& -\pi^2-\hbar q B\cdot \Sigma \notag \\
     &=&- \pi^2+2q B\cdot S. \label{I term}
\end{eqnarray}
By substituting this result back into Eq. \eqref{Pauli Eq. before manipulation, I terms}, the classic Pauli equation \cite{Pauli1927} is recovered at zeroth order in $\ell_0$ (albeit with $16$-dimensional spin matrices contained in $S$):
\begin{eqnarray}
    i \hbar \frac{\partial}{\partial t} \xi&=&\left( \frac{\pi^2}{2m}-g\frac{q}{2m} B\cdot S+q\phi\right) \xi+O(\ell_0), \label{Pauli Eq. zeroth order}
\end{eqnarray}
where $g=2$ is the gyromagnetic factor predicted by Dirac's equation \cite{Dirac1928,dirac1981principles}. Note that $S_k=(i\hbar/4)\varepsilon_{ijk}[G_i,G_j]$, $k=1,2,3$, is a 3-vector of Hermitian 16-by-16 matrices with 8-fold degenerate eigenvalues $\pm \hbar/2$, analogous to the vector of spin matrices appearing in the classic Pauli equation.\\

\subsection*{II}
\cc{The $\ell_0$-modulated term $\mathrm{II}$ on the right-hand side of Eq. \eqref{Different terms} is now examined. Noting that all indices in this term are dummy indices and using Eq. \eqref{identity 2}, one can write
\begin{widetext}
    \begin{eqnarray}
   \hspace{-.5cm} \mathrm{II}&=&\frac{\ell_0}{\hbar}\pi_k  \left(M^+_{kl}\pi_l\pi_m V_m+ V_k \pi_m M^-_{mn}\pi_n\right)\nonumber\\
    &=&\frac{\ell_0}{\hbar}\left(\pi_k \pi_l\pi_m  M^+_{kl} V_m+ \pi_m \pi_k\pi_l V_m  M^-_{kl}\right)\nonumber\\
    &=&\frac{\ell_0}{\hbar}\left[\pi_k \pi_l\pi_m  \underbrace{\left(M^+_{kl} V_m+ V_m  M^-_{kl}\right)}_{=0}+[\pi_m,\pi_k \pi_l] V_m  M^-_{kl}\right].\nonumber\\
    \label{intmd}
\end{eqnarray}
\end{widetext}
The commutator in this equation is resolved as follows:
    \begin{eqnarray}
    [\pi_m,\pi_k \pi_l]&=&[\pi_m, \pi_k] \pi_l + \pi_k [\pi_m, \pi_l]\nonumber\\
    &=& i\hbar q\left(\varepsilon_{mkj} B_j \pi_l  +\varepsilon_{mlj} \pi_k  B_j\right) .  \label{magb}
\end{eqnarray}}\\

\cc{Combining Eqs. \eqref{intmd} and \eqref{magb} yields
\begin{eqnarray}
    \mathrm{II}&=&\frac{\ell_0}{\hbar}[\pi_m,\pi_k \pi_l] V_m  M^-_{kl}\\
    &=&i q \ell_0 \left(\varepsilon_{mkj} B_j \pi_l  +\varepsilon_{mlj} \pi_k  B_j\right)V_m  M^-_{kl}\nonumber \\
    &=&i q \ell_0  \left(\varepsilon_{mkj} B_j \pi_l  +\varepsilon_{mkj} \pi_l B_j\right)V_m  M^-_{kl}\nonumber\\
    &=&i q \ell_0  \varepsilon_{mkj} \{B_j,\pi_l\}V_m  M^-_{kl}. \label{II term}
\end{eqnarray}}
Defining the second-order tensors $\mathcal{B}_{jl}=\{B_j,\pi_l\}$ and $\mathcal{V}_{jl}=\varepsilon_{mkj} V_m  M^-_{kl}$, the above expression can be simplified as follows:
\begin{eqnarray}
    \mathrm{II}=i q \ell_0 \mathcal{V}:\mathcal{B}(\pi). \label{eq86ref}
\end{eqnarray}
\subsection*{III}
Next, the third term $\mathrm{III}$ on the right-hand side of Eq. \eqref{Different terms} is considered. First, note that one can deduce from Eq. \eqref{gacom} that
\begin{eqnarray}
   \{M^+_{kl},M^-_{mn}\}&=&-2\delta_{km}\delta_{ln}+2\left(\delta_{k2}\delta_{l2}M^-_{mn}-M^+_{kl}\delta_{m2}\delta_{n2}\right)\notag \\
  &&+4\delta_{k2}\delta_{l2}\delta_{m2}\delta_{n2}
\end{eqnarray}
Let us now write:
\begin{widetext}
\begin{eqnarray}
\mathrm{III}&=&\left(\frac{\ell_0}{\hbar}\right)^2\pi_k\pi_l \pi_m \pi_n M^+_{kl} M^-_{mn}\notag \\
    &=&\frac{1}{4}\left(\frac{\ell_0}{\hbar}\right)^2\{\pi_k \pi_l, \pi_m \pi_n\}(M_{kl}^+ M_{mn}^-+M_{mn}^- M_{kl}^+)+\frac{1}{4}\left(\frac{\ell_0}{\hbar}\right)^2[\pi_k\pi_l,\pi_m \pi_n](M_{kl}^+ M_{mn}^--M_{mn}^- M_{kl}^+)\notag \\
   &=&-\underbrace{\frac{1}{2}\left(\frac{\ell_0}{\hbar}\right)^2\{\pi_k \pi_l, \pi_m \pi_n\}\delta_{km}\delta_{ln}}_{=\mathrm{IIIa}}+\underbrace{\frac{1}{2}\left(\frac{\ell_0}{\hbar}\right)^2\{\pi_k \pi_l, \pi_m \pi_n\}\left(\delta_{k2}\delta_{l2}M^-_{mn}-M^+_{kl}\delta_{m2}\delta_{n2}+2\delta_{k2}\delta_{l2}\delta_{m2}\delta_{n2}\right)}_{=\mathrm{IIIb}}\notag \\
   && +\underbrace{\frac{1}{4}\left(\frac{\ell_0}{\hbar}\right)^2[\pi_k\pi_l,\pi_m \pi_n](M_{kl}^+ M_{mn}^--M_{mn}^- M_{kl}^+)}_{=\mathrm{IIIc}}\notag \\
   &=& -\mathrm{IIIa}+\mathrm{IIIb}+\mathrm{IIIc}.\label{eq6}
\end{eqnarray}
\end{widetext}
Substituting this expression into Eq. \eqref{Pauli Eq. before manipulation, I terms} yields
\begin{eqnarray}
    i \hbar \frac{\partial \xi}{\partial t}&=&-\dfrac{\mathrm{I}-\mathrm{II}-\mathrm{IIIa}+\mathrm{IIIb}+\mathrm{IIIc}}{2m}\xi+q\phi \xi.\label{auxeq 2}
\end{eqnarray}

One can readily simplify the first symmetric term in Eq. \eqref{eq6}:\\
\begin{eqnarray}
  \mathrm{IIIa}&=& \frac{1}{2}\left(\frac{\ell_0}{\hbar}\right)^2\{\pi_k \pi_l, \pi_m \pi_n\}\delta_{km}\delta_{ln}\notag\\
   &=&\frac{1}{2}\left(\frac{\ell_0}{\hbar}\right)^2\{ \pi_k \pi_l ,\pi_k\pi_l\}\notag \\
    &=&\left(\frac{\ell_0}{\hbar}\right)^2\pi_k \pi_l  \pi_k  \pi_l \notag \\
    &=& \left(\frac{\ell_0}{\hbar}\right)^2 \pi^4. \label{IIIa term}
\end{eqnarray}
The following term is also rewritten to show that it vanishes identically:
\begin{widetext}
\begin{eqnarray}
    \mathrm{IIIb}&=&\frac{1}{2} \left(\frac{\ell_0}{\hbar}\right)^2\{\pi_k \pi_l, \pi_m \pi_n\}\left(\delta_{k2}\delta_{l2}M^-_{mn}-M^+_{kl}\delta_{m2}\delta_{n2}+2\delta_{k2}\delta_{l2}\delta_{m2}\delta_{n2}\right)\nonumber \\
    &=&\frac{1}{2} \left(\frac{\ell_0}{\hbar}\right)^2\left(\{\pi_2^2, \pi_m \pi_n\}M^-_{mn}-M^+_{kl}\{\pi_k \pi_l, \pi_2^2\}+ 2\{\pi_2^2, \pi_2^2\} \right)\nonumber\\
    &=&\frac{1}{2}\left(\frac{\ell_0}{\hbar}\right)^2 \left(\{\pi_2^2, \pi_m \pi_n\}M^-_{mn}-M^+_{kl}\{\pi_2^2,\pi_k \pi_l\}+4\pi_2^4\right)\nonumber\\
    &=&\frac{1}{2}\left(\frac{\ell_0}{\hbar}\right)^2 \left(\{\pi_2^2, \pi_k \pi_l\}\left(M^-_{kl}-M^+_{kl}\right)+4\pi_2^4\right) \nonumber\\
     &=&\frac{1}{2}\left(\frac{\ell_0}{\hbar}\right)^2\left( -2\{\pi_2^2, \pi_2^2\}+4\pi_2^4\right) \nonumber\\
     &=&0. \label{IIIb term}
\end{eqnarray}
\end{widetext}
The $\pi$-commutator in Eq. \eqref{eq6} can generally be rewritten as follows:
\begin{eqnarray}
    [\pi_k\pi_l,\pi_m \pi_n]&=&\pi_k [\pi_l , \pi_m] \pi_n +[\pi_k,\pi_m]\pi_l\pi_n \notag \\
&&+\pi_m\pi_k[\pi_l,\pi_n]+\pi_m[\pi_k,\pi_n]\pi_l.\label{xx}
\end{eqnarray}
If this term is contracted with $M^+_{kl} M^-_{mn}-M^-_{mn} M^+_{kl}$, because the tensors $M^\pm$ are symmetric, one can exchange $k$ with $l$ and $m$ with $n$. This procedure leads to 
\begin{widetext}
\begin{eqnarray}
      &&[\pi_k\pi_l,\pi_m \pi_n]\left(M^+_{kl} M^-_{mn}-M^-_{mn} M^+_{kl}\right)\notag\\
      &&=\left(\pi_k [\pi_l , \pi_m] \pi_n +[\pi_k,\pi_m]\pi_l\pi_n+\pi_m\pi_k[\pi_l,\pi_n]+\pi_m[\pi_k,\pi_n]\pi_l\right)\left(M^+_{kl} M^-_{mn}-M^-_{mn} M^+_{kl}\right)\\
      &&=\left(\{\pi_k,[\pi_l,\pi_m]\}\pi_n + \pi_m \{\pi_k,[\pi_l,\pi_n]\}\right)(M^+_{kl} M^-_{mn}-M^-_{mn} M^+_{kl})\notag\\
      &&=\{\pi_m,\{\pi_k,[\pi_l,\pi_n]\}\}(M^+_{kl} M^-_{mn}-M^-_{mn} M^+_{kl}).
\end{eqnarray}
\end{widetext}

In summary, one has
\begin{widetext}
    \begin{eqnarray}
   \hspace{-0.3cm}\mathrm{IIIc} &=&\frac{1}{4}\left(\frac{\ell_0}{\hbar}\right)^2[\pi_k\pi_l,\pi_m \pi_n]\left(M^+_{kl} M^-_{mn}-M^-_{mn} M^+_{kl}\right) \notag \\
    &=&\frac{1}{4}\left(\frac{\ell_0}{\hbar}\right)^2\{\pi_m,\{\pi_k,[\pi_l,\pi_n]\}\}\left(M^+_{kl} M^-_{mn}-M^-_{mn} M^+_{kl}\right) \nonumber \\
    &=&\frac{i\hbar q}{4}\left(\frac{\ell_0}{\hbar}\right)^2\{\pi_m,\{\pi_k,\varepsilon_{lnj} B_j\}\}\left(M^+_{kl} M^-_{mn}-M^-_{mn} M^+_{kl}\right). \label{eqr2}
\end{eqnarray}
Defining the fourth-order tensors $\mathcal{F}_{klmn}=\{\pi_m,\{\pi_k,\varepsilon_{lnj} B_j\}\}$ and $\mathcal{M}_{klmn}=M^+_{kl} M^-_{mn}-M^-_{mn} M^+_{kl}$, the above expression can be written as
\begin{eqnarray}
    \mathrm{IIIc}=\frac{i\hbar q}{4}\left(\frac{\ell_0}{\hbar}\right)^2 \mathcal{M}:\mathcal{F}(\pi). \label{IIIc term}
\end{eqnarray}
\end{widetext}
\cc{After substituting Eqs. \eqref{I term}, \eqref{eq86ref}, \eqref{IIIa term}, \eqref{IIIb term}, and \eqref{IIIc term} into Eq. \eqref{auxeq 2}, the non-relativistic limit of Eq. \eqref{Symbolic 32d (min. coupling)} is obtained: 
\begin{widetext}
    \begin{eqnarray}
 \hspace{-0.3cm}  i \hbar \frac{\partial}{\partial t} \xi&=&\bigg[ \underbrace{\frac{\pi^2}{2m} -g\frac{q}{2m} S\cdot B}_{-\mathrm{I}/(2m)} +q\phi+\underbrace{\frac{i q \ell_0}{2m} \mathcal{V}:\mathcal{B}(\pi)}_{\mathrm{II}/(2m)} +\underbrace{\left(\frac{\ell_0}{\hbar}\right)^2\frac{ \pi^4}{2m}}_{=\mathrm{IIIa}/(2m)}-\underbrace{\frac{i\hbar q}{8m}\left(\frac{\ell_0}{\hbar}\right)^2\mathcal{M}:\mathcal{F}(\pi)}_{\mathrm{IIIc}/(2m)}\bigg] \xi.\label{Pauli Eq. final}
\end{eqnarray}
\end{widetext}}

This equation is the general expression for the $\ell_0$-analogue of the Pauli equation corresponding to the analogue Dirac equation \eqref{Symbolic 32d (min. coupling)}.

\section{Example \label{example sec}}
\begin{figure}[t!]
\begin{psfrags}
    \psfrag{x}{\hspace{0.12cm}\footnotesize$x$}
    \psfrag{y}{\hspace{0.03cm}\footnotesize$y$}
    \psfrag{z}{\hspace{-0.03cm}\footnotesize$z$}
    \psfrag{h}{\footnotesize$\sim \beta_3$}
        \psfrag{g}{\hspace{0.5cm}\footnotesize$B$}
        \psfrag{e}{\footnotesize$\mathcal{S}$}
        \psfrag{f}{\footnotesize$\mathcal{N}$}
    \psfrag{d}{atom beam \hspace{1cm}$\partial_z B=(\beta_1+\beta_3 x^2) e_z$}
        \psfrag{i}{magnetic field}
    \centering
\includegraphics[width=0.85\linewidth]{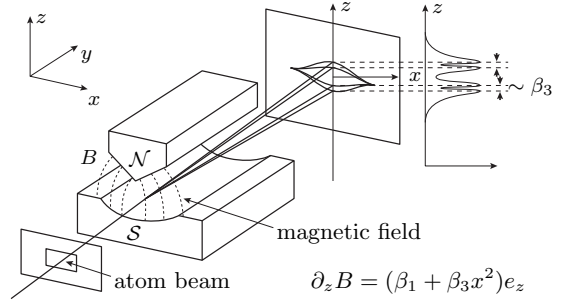}
\end{psfrags}
    \caption{{The $\ell_0$-analogue of the Pauli equation derived from the modified Klein--Gordon equation \eqref{Mindlin Pavlopoulos equation}} predicts that in a Stern--Gerlach experiment, hyperfine splitting in the detected beam intensity occurs \cc{under a symmetric higher-order inhomogeneity in the magnetic field}.}
    \label{Figure 1}
\end{figure}

In the classic Stern--Gerlach experiment, a collimated beam of silver atoms, each possessing a single unpaired electron, is sent through an inhomogeneous magnetic field, which deflects the beam in positive or negative $z$-direction, depending on the electrons' spins \cite{stern1922}. An illustration is given in Fig. \eqref{Figure 1}. \footnote{Shown in Fig. \ref{Figure 1} is an adaptation of the commemorative plaque of the Stern--Gerlach experiment at the Physikalischer Verein in Frankfurt.}  

  A theoretical model of a Stern--Gerlach experiment \cite{rosen1932double, sidles1992folded, batelaan1997stern, hermanspahn2000observation, li2007generalized, japha2023quantum} is now considered in the context of Eq. \eqref{Pauli Eq. final}, which is assumed to be valid in this example. Furthermore, a number of assumptions are made which allow for simplification of that equation:
\begin{enumerate}[label=\textbf{A\arabic*}]
\item The electron beam passes through the origin $x=y=z=0$ and remains at the middle line $x=0$ throughout the process.
    \item A solenoidal magnetic field with a generic higher-order inhomogeneity with characteristic dimension $d=\sqrt{|\beta_1/\beta_3|}$ is assumed: \begin{eqnarray}
       B= \begin{pmatrix}
        -\beta_1 x -\dfrac{\beta_3 x^3}{3}\\
        0\\
        B_0 +\beta_1 z + \beta_3 x^2 z
    \end{pmatrix}. \label{magnetic field}
    \end{eqnarray}
    
    \item The vector potential $A$ corresponding to $B=\nabla\times A$ is given by \begin{eqnarray}
       A= \begin{pmatrix}
        0 \\
        B_0 x+\beta_1 xz + \beta_3 \dfrac{x^3}{3} z \\
        0     \end{pmatrix}. \label{A potential}
    \end{eqnarray}
    The entries of $A$ are identically zero at all times along the electron beam.
    
    \item All momentum terms, i.e., terms where $\pi$ is left-multiplied with the wave function $\xi$, are neglected and only the terms involving products of the magnetic field $B$ (or its derivatives) with $\xi$ are considered. 

     \item All precession effects and the scalar electromagnetic potential $\phi$ are neglected.
     
    \item To model the deflection of the electron beam by the magnetic field, the simplified treatment of \cite[Sec. 9.1]{ballentine2014quantum} is followed. For this approach, Eq. \eqref{Pauli Eq. final} is rewritten as \begin{eqnarray}
        i \hbar \frac{\partial}{\partial t} \xi=H\xi,
    \end{eqnarray}
    where $H$ is the Hamiltonian operator. While passing through the magnetic field $B$, the wavefunction evolves like $\xi\sim e^{-i H t/\hbar}$, such that, after a time $\Delta t$ spent inside the magnetic field, it will have acquired the impulse $p\xi=-\Delta t \nabla H  \xi$. According to this model, the vertical deflection of the beam is governed by the $z$-component of the dipole force operator $F=-\nabla H$. \\
\end{enumerate}

\cc{The simplifying assumptions $\textbf{A4}$ to $\textbf{A6}$ serve to more clearly illustrate the main results and follow similar assumptions made in textbooks \cite{ballentine2014quantum,griffiths_qm}. A more detailed analysis of spin precession \cite{larmor1897,thomas1926} in the context of the higher-order electrodynamic framework considered here is left for future research.}

The linear, $\beta_1$-modulated inhomogeneities in Eq. \eqref{magnetic field} are also found in classic textbook models \cite{ballentine2014quantum,griffiths_qm}, while an additional cubic term in the $z$-component of the magnetic field with coefficient $\beta_3$ is considered here. {Motivated by the symmetric geometry of the classic Stern--Gerlach experiment (see Fig. \eqref{Figure 1}), the higher-order term is assumed to be symmetric in $x$. This nonlinearity does not change the prediction of the classic Pauli equation approach found in textbooks, but does come to bearing here due to the presence of higher-order derivatives in Eq. \eqref{Pauli Eq. final}. The terms in the $x$-component in Eq. \eqref{magnetic field} were chosen to balance the terms in the $z$-component when taking the divergence of $B$, ensuring a solenoidal magnetic field.} 

\cc{Due to assumptions $\textbf{A1}$ to $\textbf{A6}$, the terms $\mathrm{II}$ and $\mathrm{IIIa}$ in Eq. \eqref{Pauli Eq. final} do not contribute to the beam deflection in $z$-direction, and the term $\mathrm{I}$ simplifies to just the magnetic dipole:
\begin{eqnarray}
    \mathrm{I}\rightarrow g q\, S\cdot B .
\end{eqnarray}}
A crucial contribution comes from the term $\mathrm{IIIc}$, which is now examined by simplifying Eq. \eqref{eqr2}. In particular, the following simplifications apply to the term $\{\pi_m,\{\pi_k,\left[\pi_l,\pi_n\right]\}\}=i\hbar q\{\pi_m,\{\pi_k,\varepsilon_{lnj} B_j\}\}$:
\begin{eqnarray} 
\{\pi_k,\left[\pi_l,\pi_n\right]\}&\rightarrow&\hbar^2 q\,\varepsilon_{lnj}B_{j,k},\\
\{\pi_m,\{\pi_k,\left[\pi_l,\pi_n\right]\}\}&\rightarrow &-i\hbar^3 q\,\varepsilon_{lnj}B_{j,km}. \label{eqx}
\end{eqnarray}
Now,  using the simplified notation $[klmn]=M^+_{kl} M^-_{mn}-M^-_{mn} M^+_{kl}$, the expression $\mathrm{IIIc}$ can be reformulated as
\begin{eqnarray}
   &&\mathrm{IIIc} =\frac{1}{4}\left(\frac{\ell_0}{\hbar}\right)^2[\pi_k\pi_l,\pi_m \pi_n]\left(M^+_{kl} M^-_{mn}-M^-_{mn} M^+_{kl}\right) \notag \\
    &=&\frac{1}{4}\left(\frac{\ell_0}{\hbar}\right)^2\{\pi_m,\{\pi_k,[\pi_l,\pi_n]\}\}\left(M^+_{kl} M^-_{mn}-M^-_{mn} M^+_{kl}\right) \notag\\
    &=&-\left(\frac{\ell_0}{\hbar}\right)^2\frac{i\hbar^3 q}{4}\varepsilon_{lnj}B_{j,km}\left(M^+_{kl} M^-_{mn}-M^-_{mn} M^+_{kl}\right) \notag\\
    &=&-\left(\frac{\ell_0}{\hbar}\right)^2\frac{i\hbar^3 q}{4}\varepsilon_{lnj}B_{j,km}[klmn] \notag \\
    &=&-\left(\frac{\ell_0}{\hbar}\right)^2\frac{i{\sqrt{2}}\hbar^3 q}{4}\Theta_j B_j \notag \\
    &=&-\left(\frac{\ell_0}{\hbar}\right)^2\frac{i{\sqrt{2}}\hbar^3 q}{4}\Theta\cdot B\notag \\
    &=&-{\sqrt{2}}\ell_0^2 q\,\mathcal{T}\cdot B .\label{eqr}
\end{eqnarray}
In Eq. \eqref{eqr}, the following differential operators were defined:
\begin{eqnarray}
    \Theta_j&=&\frac{1}{{\sqrt{2}}}\varepsilon_{lnj} [klmn] \frac{\partial}{\partial k}\frac{\partial}{\partial m}, \label{equ}\\
    \mathcal{T}&=&\frac{i\hbar}{4} \Theta.
\end{eqnarray}
The individual entries of the operator $\Theta$ are
\begin{widetext}
\begin{eqnarray}
    \begin{pmatrix}
        \Theta_1\\
        \Theta_2\\
        \Theta_3
    \end{pmatrix}&=&\frac{1}{{\sqrt{2}}}\begin{pmatrix}
[ k2m3 ] - [ k3m2 ] \\
[ k3m1 ] - [ k1m3 ] \\
[ k1m2 ] - [ k2m1 ]
\end{pmatrix}\frac{\partial}{\partial k}\frac{\partial}{\partial m}\notag \\
    &=&\frac{1}{\sqrt{2}} \begin{pmatrix}
\left([1213] - [1312]\right)\partial_x^2 + \left([1223] - \cancel{[1322]}\right) \partial_x \partial_y + \left([1233] - [1332]\right) \partial_x \partial_z \\
\quad + \left(\cancel{[2213]} - [2312]\right) \partial_y \partial_x + \left(\cancel{[2223]} - \cancel{[2322]}\right)\partial_y^2 + \left(\cancel{[2233]} - [2332]\right) \partial_y \partial_z \\
\quad + \left([3213] - [3312]\right) \partial_z \partial_x + \left([3223] - \cancel{[3322]}\right) \partial_z \partial_y + \left([3233] - [3332]\right)\partial_z^2 \\[1.2em]

\left([1311] - [1113]\right)\partial_x^2 + \left([1321] - [1123]\right) \partial_x \partial_y + \left([1331] - [1133]\right) \partial_x \partial_z \\
\quad + \left([2311] - [2113]\right) \partial_y \partial_x + \left([2321] - [2123]\right)\partial_y^2 + \left([2331] - [2133]\right) \partial_y \partial_z \\
\quad + \left([3311] - [3113]\right) \partial_z \partial_x + \left([3321] - [3123]\right) \partial_z \partial_y + \left([3331] - [3133]\right)\partial_z^2 \\[1.2em]

\left([1112] - [1211]\right)\partial_x^2 + \left(\cancel{[1122]} - [1221]\right) \partial_x \partial_y + \left([1132] - [1231]\right) \partial_x \partial_z \\
\quad + \left([2112] - \cancel{[2211]}\right) \partial_y \partial_x + \left(\cancel{[2122]} - \cancel{[2221]}\right)\partial_y^2 + \left([2132] - \cancel{[2231]}\right) \partial_y \partial_z \\
\quad + \left([3112] - [3211]\right) \partial_z \partial_x + \left(\cancel{[3122]} - [3221]\right) \partial_z \partial_y + \left([3132] - [3231]\right)\partial_z^2
\end{pmatrix}. \label{Theta operator written out}
\end{eqnarray}
\end{widetext}
where the shorthand notation $\partial_k=\partial/\partial_k$ was used, with $k,m\in\{1,2,3\}$ corresponding to $(x,y,z)$. The crossed out terms in \eqref{Theta operator written out} vanish because of the commutation property of the matrix $G_5=I_{16}$. As evident in Eq. \eqref{Theta operator written out}, there is a preferred direction ($y$) in this model, which is here assumed to be aligned with the beam axis, see Fig. \ref{Figure 1}.

The simplified $\ell_0$-analogue of the Pauli equation \eqref{Pauli Eq. final} can now be written as
\begin{eqnarray}
        i \hbar \frac{\partial}{\partial t} \xi=\underbrace{-\left[g\frac{q}{2m}\left( S- \frac{ \ell_0^2 }{{\sqrt{2}}} \mathcal{T}\right) \cdot B \right]}_{H}\xi. \label{simplified Pauli eq}\\
        \nonumber
    \end{eqnarray}
    
\cc{By the form of the magnetic field assumed in $\textbf{A2}$, the only terms in $\mathcal{T}$ that could contribute to the $z$-component of $-\nabla H$ along the beam are the second-order derivatives in $x$:
    \begin{eqnarray}
   \mathcal{T}\rightarrow \frac{i\hbar}{4{\sqrt{2}}}\begin{pmatrix}
 [G_7,G_8]\partial_x^2 \\
 * \\
{\sqrt{2}}[G_4,G_7]\partial_x^2 
\end{pmatrix}.
\end{eqnarray}}
The Hamiltonian $H$ defined through Eq. \eqref{simplified Pauli eq} can be written out as follows:
\begin{widetext}
\begin{eqnarray}
    H&=&-\left[g\frac{q}{2m}\left( S- \frac{ \ell_0^2 }{{\sqrt{2}}} \mathcal{T}\right) \cdot B \right]\notag\\
    &=&-\left[g\frac{q}{2m}\left( S- \frac{ \ell_0^2 }{{\sqrt{2}}} \mathcal{T}\right) \cdot \begin{pmatrix}
        -\beta_1 x -\dfrac{\beta_3 x^3}{3}\\
        0\\
        B_0 +\beta_1 z + \beta_3 x^2 z
    \end{pmatrix} \right]\notag\\
    &=&-g\frac{q}{2m}\left[S_1 \left(-\beta_1 x -\frac{\beta_3 x^3}{3}\right)+\frac{i\hbar\ell_0^2 \beta_3 x}{4} [G_7,G_8] +S_3\left(B_0 +\beta_1 z + \beta_3 x^2 z\right)-\underbrace{\frac{i\hbar \ell_0^2 \beta_3 z}{2{\sqrt{2}}}[G_4,G_7] }_{={\sqrt{2}}\ell_0^2\beta_3 z T_3}\right].
\end{eqnarray}
\end{widetext}
The $z$-component of the dipole force operator $F=-\nabla H$ has full rank and is given by 
\begin{eqnarray}
    F_z=\mu\left[\left(\beta_1 +\beta_3 x^2\right)S_3-{\sqrt{2}}\beta_3 \ell_0^2 T_3\right],
\end{eqnarray}
where $\mu=gq/(2m)$ and $T_3=(i\hbar/4)[G_4,G_7]$. At the location of the beam ($x=0$), by Eqs. \eqref{ev1} to \eqref{ev4}, the 16 eigenvalues of $F_3$ are given by the following fourfold degenerate set (recall that $d=\sqrt{|\beta_1/\beta_3|}$):
\begin{eqnarray}
    \lambda_{++}&=&\frac{\mu\beta_1 \hbar}{2}\left(1+{\sqrt{2}}\frac{\ell_0^2}{d^2}\right), \label{l1}\\
    \lambda_{+-}&=&\frac{\mu\beta_1 \hbar}{2}\left(1-{\sqrt{2}}\frac{\ell_0^2}{d^2}\right),\\
    \lambda_{-+}&=&-\frac{\mu\beta_1 \hbar}{2}\left(1-{\sqrt{2}}\frac{\ell_0^2}{d^2}\right),\\
    \lambda_{--}&=&-\frac{\mu\beta_1 \hbar}{2}\left(1+{\sqrt{2}}\frac{\ell_0^2}{d^2}\right). \label{lend}
\end{eqnarray}
The corresponding eigenvectors can be computed using the supplemental code \cite{supplement}. Depending on the alignment of the wavefunction $\xi$ with the eigenspaces of $F_z$, the electrons will accumulate positive or negative momentum in $z$-direction. Statistically, these positive and negative portions will be each evenly split around the mean values $(\lambda_{\pm +}+\lambda_{\pm -})/2$, which correspond to the classic textbook values and are independent of $\ell_0$ and $\beta_3$. See Fig. \ref{Figure 1} for an illustration of this result. Such hyperfine splitting can only be expected to be clearly observed if $d$ is of the same order of magnitude as $\ell_0$ and if the time $\Delta t$ spent by the beam in the magnetic field is sufficiently long. {Had one not assumed that the beam is aligned with the $y$-axis, a similar fine-splitting as predicted by Eqs. \eqref{l1} to \eqref{lend} would result, albeit with different numerical factors (depending on the specific alignment with the coordinate axes) before the $\ell_0^2$-modulated terms.}
\\

\section{Discussion}
This study has extended the analogy used in \cite{Pavlopoulos67} of replacing the d'Alembert operator by a higher-order operator to the Klein--Gordon, Dirac and Pauli equations. Thereby, a \cc{universal} length scale $\ell_0$ is introduced, which leads to a modified framework of relativistic electrodynamics, in which the Lorentz symmetry is replaced by a larger symmetry, which, so far, remains unknown. It was shown that, \cc{under standard assumptions}, the $\ell_0$-analogue of the Pauli equation derived in this manner predicts hyperfine splitting in the Stern--Gerlach experiment under \cc{a symmetric higher-order inhomogeneity} in the magnetic field. 

The models presented here could be used to describe quantum phenomena involving the stable occupation number $16$, which have been observed in nuclear science \cite{ozawa2000} and organometallic chemistry \cite{tolman1972,barry2013thermochromic}. Note that the number $16$ is not naturally distinguished in standard quantum mechanics (see \cite{Gies2024} for an example of a nonstandard theory in which it is), but arises generically in the context of Eq. \eqref{simplified Pauli eq}, which, under a symmetrically inhomogeneous magnetic field, predicts four different values of spin, which are more distinct for stronger inhomogeneities. This fourfold multiplicity of spin implies a secondary series of stable occupation numbers (besides $2,8,\dots$) starting with the numbers $4$ and $16$.

\cc{An even finer splitting of the energy levels can be expected if the $\mathrm{II}$ term in Eq. \eqref{Pauli Eq. final} contributes. This could occur, for example, if an asymmetric nonlinearity proportional to $xz$ were included in the magnetic field \eqref{magnetic field}. Interestingly, if $\ell_0$ were taken as the classical electron radius as proposed in \cite{Pavlopoulos67} and if $|\pi|\approx \hbar \lambda_\text{C}^{-1}$, where $\lambda_\text{C}=\hbar/(mc)$ is the reduced Compton wavelength, then the contribution of the $\mathrm{II}$ term would be of the order of the fine structure constant $\alpha=\ell_0/\lambda_\text{C}$. Similarly, the contribution of $\mathrm{III}$ would be of the order of $\alpha^2$.}

{It is also mentioned here that Eq. \eqref{RHS} features two different terms, $M^+$ and $M^-$, in the particle and antiparticle wave equations, respectively. This asymmetry is absent in the terms up to order $O(\ell_0)$, which, as shown in Eq. \eqref{Pauli Eq. zeroth order}, recover the classic Pauli equation. Besides this inherent particle-antiparticle asymmetry, another curious feature derived from Eq. \eqref{Mindlin Pavlopoulos equation} is the fact that space is oriented, i.e., there is always one preferred direction at each point (see, for example, Eq. \eqref{Theta operator written out}). This feature relates to the last paragraph of \cite{Pavlopoulos67}, wherein the need for extending special relativity into the realm of micropolar mechanics \cite{cosserat1909theorie} was expressed. To date, the development of such a program remains an open task.} 

\subsection{Application of the method to other models}
{Higher-order wave equations arise in a wide range of physical contexts from shallow water waves \cite{Korteweg1895,Zabusky1965,Hirota1971,Fokas1981,Goldstein1991}, nonlocal field theories \cite{bopp1940,Lande1940a,Lande1940b,Lande1944,Podolsky1942,Podolsky1944,Montgomery1947,Green1947,PodolskySchwed1948,deBroglie1949,Pavlopoulos67,Ji2019}, and doubly special relativity \cite{Amelino-Camelia2002,Amelino-Camelia2002a,Magueijo2002,Judes:2002bw,KowalskiGlikman:2004qa,Liberati:2004ju,Jizba:2012,relancio2024doubly}, to elasticity \cite{Euler1744,Love1888,Timoshenko1921,Mindlin1951,Reddy1979,LandauElasticity,Pocivavsek2008Science}, micropolar media \cite{cosserat1909theorie,Toupin1962,Mindlin1962,Truesdell1965,Eringen1999,Merkel2011,Xu2020,Nassar2020,Kishine2020}, and topological meta-materials \cite{Fan2019Elastic,Rosa2019EdgeStates,Deymier2022Revealing}. Despite their ubiquity, the polarization dynamics and symmetries of these models remain poorly understood, and the ansatz \eqref{ansatz 1} is insufficient to capture the higher-order derivatives that appear in their governing equations.}

{Beyond its primary focus on electromagnetic phenomena, this work seeks to extend the algebraic factorization method underlying the derivation of the Dirac equation to higher-order equations and to make their spin dynamics accessible to systematic analysis. To avoid overshadowing the simplicity of this method, it is here applied to the specific example of Eq. \eqref{Mindlin Pavlopoulos equation}, and a more general exposition is postponed to future research. Nevertheless, to exemplify the applicability of the method to another well-known example of a higher-order equation, in appendix \ref{appendix}, the first steps of the algebraic factorization procedure are reproduced for the equation of Bopp \cite{bopp1940}, given by $(\Box - \kappa_0^2)\Box\psi = 0$, where the reciprocal $\kappa_0^{-1}$ is a \cc{universal} length scale similar to $\ell_0$ in Eq. \eqref{Mindlin Pavlopoulos equation}.}

\section{Conclusions}
\cc{The concept of a universal length scale has been a fascination among physicists at least since Heisenberg's 1938 review article \cite{Heisenberg1938} and remains both a puzzling and promising research area until today \cite{Hossenfelder2013}. This article has sought to make a contribution to this field by proposing a Dirac-like quantum electrodynamic theory based on a modified wave operator, wherein Lorentz symmetry is explicitly broken by the length scale $\ell_0$. The full Lie group analysis of the operator fourth-order operator \eqref{higher order operator} is deferred to a later point due its complexity. This will also shed light on the micropolar spacetime induced by \eqref{Mindlin Pavlopoulos equation}. Additionally, an alternative approach based on a simpler, more accessible transformation theory also involving a \cc{universal} length scale could be pursued in the future.}

\section*{Acknowledgements}
The work of T.P. was supported by the Swiss National Science Foundation under Grant Agreement No. 225619.

\section*{Authors' contribution statement}
T. P. wrote the initial draft of the manuscript, including the algebraic factorization of the modified Klein--Gordon equation and non-relativistic limit derivation. F. K. led the study of the broken Lorentz symmetry for non-zero values of $\ell_0$. Both authors contributed to the revision of the manuscript.

\bibliography{apssamp}
\clearpage 
\appendix
\section{Factorization of Bopp's equation \label{appendix}}
{The equation of Bopp is given by $(\Box - \kappa_0^2)\Box\psi = 0$, or, written out,
\begin{widetext}
\begin{eqnarray}
    \left( \left[ \frac{\partial^2}{\partial x^2}
+ \frac{\partial^2}{\partial y^2}
+ \frac{\partial^2}{\partial z^2} -\frac{1}{c^2}\frac{\partial^2}{\partial t^2}\right] - \kappa_0^2 \right)
\left[ \frac{\partial^2}{\partial x^2}
+ \frac{\partial^2}{\partial y^2}
+ \frac{\partial^2}{\partial z^2} -\frac{1}{c^2}\frac{\partial^2}{\partial t^2}
\right]\psi = 0
\end{eqnarray}
\end{widetext}
To factorize this equation in the sense that 
\begin{eqnarray}
    \mathcal{D}_0^2\Psi=(\Box-\kappa_0^2)\Box\Psi,
\end{eqnarray} 
one could make the ansatz
\begin{widetext}
\begin{eqnarray}
\mathcal{D}_0&=& i\kappa_0\left(\Gamma_0 \frac{i}{c}\frac{\partial}{\partial t}+\Gamma_1 \frac{\partial}{\partial x}+\Gamma_2 \frac{\partial}{\partial y} +\Gamma_3 \frac{\partial}{\partial z} \right)+ \Gamma_4\frac{1}{c^2}\frac{\partial^2}{\partial t^2}+ \Gamma_5\frac{\partial^2}{\partial x^2}+\Gamma_6\frac{\partial^2}{\partial y^2}+\Gamma_7\frac{\partial^2}{\partial z^2}\notag\\
&&+\frac{\sqrt{2}i}{c}\left(\Gamma_8\frac{\partial}{\partial t}\frac{\partial}{\partial x}+\Gamma_9\frac{\partial}{\partial t}\frac{\partial}{\partial y}+\Gamma_{10}\frac{\partial}{\partial t}\frac{\partial}{\partial z}\right)+{\sqrt{2}}\left(\Gamma_{11}\frac{\partial}{\partial x}\frac{\partial}{\partial y}+\Gamma_{12}\frac{\partial}{\partial x}\frac{\partial}{\partial z}+\Gamma_{13}\frac{\partial}{\partial y}\frac{\partial}{\partial z} \right).\label{ansatz bopp}
\end{eqnarray}
\end{widetext}
The dimension of these $\Gamma$ matrices and the wavefunction $\Psi$ would be $2^{14/2}=128$. }
\end{document}